\documentclass{ws-mpla}
\usepackage{amscd,amssymb,amsmath,amsfonts}

\def\susix{SU(6)}
\def\suthr{SU(3)}

\def\sutwo{SU(2)}
\def\uone{U(1)\,}
\def\uoneb{U(1)$_\mathrm{B}$}
\def\uonec{U(1)$_\mathrm{C}$}

\def\sutwol{SU(2)$_\mathrm{L}$}
\def\suthrc{SU(3)$_\mathrm{C}$}

\def\suthrh{SU(3)$_\mathrm{H}$}
\def\uoney{U(1)$_\mathrm{Y}$}
\def\M4{\mathcal{M}^4 \times (S^1/Z_2)}
\def\be{\begin{equation}}
\def\ee{\end{equation}}
\def\bea{\begin{eqnarray}}
\def\eea{\end{eqnarray}}
\def\bw{\begin{widetext}}
\def\ew{\end{widetext}}
\def\j#1#2#3#4{{\it #1} {\bf #2} #3 #4}
\def\pl{Phys. Lett.}
\def\np{Nucl. Phys.}
\def\pr{Phys. Rev.}
\def\jhep{JHEP}
\def\npps{Nucl. Phys. Proc. Suppl.}
\begin{document}

\title{NEAR-BRANE \susix-ORIGIN HIGGS IN SCHERK-SCHWARZ BREAKING OF 5-DIMENSIONAL \susix \,  GUT}

\author{A. Hartanto}
\address{Theoretical Physics Laboratory, Theoretical High Energy Physics \& Instruments (THEPI)\\ Research Division and Indonesia Center for Theoretical and Mathematical Physics (ICTMP) \\
Institut Teknologi Bandung, Jl. Ganesha 10,
Bandung 40132 Indonesia, \\
andreashartanto@cbn.net.id, espdir@cbn.net.id}

\author{F.P. Zen}
\address{Theoretical Physics Laboratory, Theoretical High Energy Physics \& Instruments (THEPI)\\ Research Division and Indonesia Center for Theoretical and Mathematical Physics (ICTMP) \\
Institut Teknologi Bandung, Jl. Ganesha 10,
Bandung 40132 Indonesia, \\
fpzen@fi.itb.ac.id}

\author{J.S. Kosasih}
\address{Theoretical Physics Laboratory, Theoretical High Energy Physics \& Instruments (THEPI)\\ Research Division and Indonesia Center for Theoretical and Mathematical Physics (ICTMP) \\
Institut Teknologi Bandung, Jl. Ganesha 10,
Bandung 40132 Indonesia, \\
jusak@fi.itb.ac.id}

\author{L.T. Handoko}
\address{Theoretical and Computational Physics Group, Research Center for Physics,  Indonesian Institute of Science, Tangerang, Banten, Indonesia\\
Department of Physics, University of Indonesia, Depok, West Java, Indonesia\\
handoko@lipi.go.id, handoko@fisika.lipi.go.id, handoko@fisika.ui.ac.id}

\maketitle

\pub{Received (Day Month Year)}{Revised (Day Month Year)}

\begin{abstract}
The symmetry breaking of 5-dimensional \susix \, GUT  is realized by Scherk-Schwarz mechanisms through trivial and pseudo non-trivial orbifold $S^1/Z_2$ breakings to produce dimensional deconstruction 5D \susix$\rightarrow$4D \susix. The latter also induces near-brane weakly-coupled \susix \, Baby Higgs to further break the symmetry into \suthrc$\otimes$\suthrh$\otimes$\uonec. The model successfully provides  a scenario of the origin of (Little) Higgs from GUT scale, produces the (intermediate and light) Higgs boson with the most preferred range  and establishes coupling unification and compactification scale correctly.

\end{abstract}

\ccode{PACS Nos.: 12.10.Dm, 12.60.Cn, 12.60.Fr, 11.15.Ex}

\section{Introduction}
In the last decades, some efforts have been dedicated
to investigate gauge theories with larger gauge symmetries
inspired by the successful electroweak theory [1].
Those theories assume gauge invariances under particular
symmetries larger than the SM's one, but contain
\suthrc$\otimes$\sutwol$\otimes$\uoney \,  as a part of its subgroups at
electroweak scale. One of them is the GUT model based
on \susix \, group [2]. The model suggests that the electroweak
scale physics is realized through breaking patterns
\susix$\rightarrow$\suthrc$\otimes$\suthrh$\otimes$\uonec, \, and subsequently
\suthrh$\rightarrow$\sutwol$\otimes$\uoneb. Unfortunately the model suffers from non-existance of appropriate Higgs multiplet [3].

Following recent progress on extra dimension physics,
non-Higgs mechanisms  have been presented  in some previous works. The so-called Scherk-Schwarz mechanisms  dynamically breaks the symmetry induced by the orbifold of extra dimension [4-6]. Instead of  directly breaking
the symmetry, the effect of compactifying the extra dimension is considered to induce the Higgs bosons itself.
This approach is known as the Higgs-gauge boson unification [7]. Recently a grand gauge-Higgs unification based on 5D \susix \, compactified on an orbifold $S^1/Z_2$ has been published with fermions in two 6-plet and one 20-plet which shows no proton decay at tree level but a little low compactification scale and heavy Higgs [20].

Little Higgs, as pseudo Nambu-Goldstone boson (PNB), can be obtained from the breaking of shift symmetry so that PNB gets the mass or global symmetry breaking has taken place [8,17-19] which in conjunction with 5D \susix \, the scalars can come from the fifth component of 5D gauge boson [6,7,20] and/or directly from the bulk [16].

In this paper, special conditions of Scherk-Schwarz mechanisms are utilized to resolve
the problem of breaking the \susix \,  GUT. The first  trivial breaking and the second non-trivial breaking pattern are realized by compactification of orbifold $S^1/Z_2$ in 5-dimensional (5D) \susix \, in parallel at the same time, not like the trivial the (pseudo) non-trivial condition generates the scalar bosons. The unperiodic 5D scalar contains the periodic 5D scalar with extra-dimensional global symmetry for small extra dimension in the so-called near-brane area [8-23]. The global symmetry \susix \, comes from 5D \susix \, gauge symmetry based on AdS/CFT correspondence [20]. Here, in the near-brane area, the first symmetry breaking of 5D \susix$\rightarrow$ 4D \susix \, is triggered by Scherk-Schwarz mechanisms and followed by trivial  and pseudo non-trivial Orbifold breaking  [10,11,15] to produce \susix-origin  Little (Baby) Higgs scalar as the origin of \susix \,  will-be-SimplestLittle  and \susix \, Baby Higgs scalars successively. Trivial Orbifold Breaking (TOB) and Pseudo non-trivial Orbifold Breaking (POB)  which perform dimensional deconstruction but still keep the symmetry intact, in principle, do produce 'exact' scalar boson, and the perturbative, incomplete series of exponential form of strongly-coupled  \susix-origin Little (Baby) Higgses. It means that \susix-origin  Little (Baby) Higgses decouple from and cannot exist in \susix \, GUT while \susix \, Baby Higgses are the perturbative zero mode with sextet containing triplet of Little(-like) Higgs. The  \susix \,  Baby Higgses  produce SM-like Higgses and become the topic of discussion in this paper. The second symmetry breaking  of 4D \susix$\rightarrow$ 4D \suthr$\otimes$\suthr$\otimes$\uone is  performed by \susix \, Little-like Higgs through orbifold-based field re-definition and the broken shift symmetry induced by the properties of VE$V$ in lower-near-brane [15,16]. The VE$V$s are obtained from two Scherk-Schwarz parameters [4-7].

One can immediately predict the birth of \suthr \, Little Higgses from the \susix-origin Little Higgses. This derivation is indeed workable and quite successful.

The paper is organized as follows, first special conditions of Scherk-Schwarz breaking, the trivial and pseudo non-trivial orbifold $S^1/Z_2$ breaking [15,22,24] are revealed in the next Section, then 5D model of \susix \, with 2 branes and the bulk [32,33] where gauge bosons and scalar bosons live in near-brane area $(y \sim 0)$ which will provide \susix-origin Little Higgs, and \susix \, Baby Higgs which is basically weakly-coupled. The two have been well reconciled within the model as well as \susix \, GUT and Baby Higgs.

The pseudo non-trivial symmetry breaking to \suthr$\otimes$\suthr$\otimes$\uone \, is explained in the next section. Subsequently it is shown that the emerging  gauge bosons from broken 5-dimensional \susix \,  could be considered as scalar boson [6,7,20] which provides the Coleman Weinberg potential for radiative symmetry breaking of 4D \susix. Before summarizing the results, a brief discussion on the order estimations of relevant physical observables within the model is given.

\section{The Scherk-Schwarz and Orbifold Breaking of \susix }
First of all let us consider the orbifold breaking in 5D \susix \, compactified in $\mathcal{M}4\times S^1/Z_2$. However before discussing the details, a brief review on Scherk-Schwarz mechanisms on orbifold $S^1/Z_2$ is given below.

\subsection{Symmetry breaking in $\mathcal{M}4\times S^1/Z_2$ through
Scherk-Schwarz mechanism}

The invariance of a theory compactified on 5-dimensional space, $\mathcal{M}4\times S^1/Z_2$, demands $\mathcal{L}_5[\phi(x, y)] =\mathcal{L}_5[\phi(x, \tau_g(y))]$. The ordinary compactification satisfies $\phi(x, \tau_g(y)) = \phi(x, y)$ which is a special case of general Scherk-Schwarz compactification condition $\phi(x, \tau_g(y)) =T_g\phi(x, y)$ [6,10,11]. Here, $\tau_g(y)$ is the operator mapping the point $y$, while $T_g$ is the twist transformation operator. Orbifold compactification has, in general, similar principles written as $\phi(x, \zeta_2(y)) = Z_2\phi(x, y)$. In case of orbifold $S^1/Z_2$ with one extra dimension, this has singularities at the fixed points after modding out $S^1$ and induces $\zeta_2(y) =-y$ which obviously satisfies the condition $\zeta^{2}_2= 1$ and $Z^{2}_2= 1$ with eigenvalues of $\pm 1$ [5,6]. This means the subspace spanned by $Z_2$ can be generated either by $Z_2 = \pm 1$ or $\sigma^3$, and written in a diagonal bases as,
\be
Z=\begin{pmatrix}
  \sigma_3  & 0 \\
  0 & \pm 1 \\
\end{pmatrix}.
\ee
Anyway, an operator $T_g$ corresponding to certain local or global symmetry and characterizing the compactification in orbifolds which satisfies consistency condition, $T_gZ_2T_g = Z_2$, can also be expressed as,
\be \label{eq:gsym}
T_g=e^{2 i\pi \vec{\beta} \cdot \vec{\lambda}}=e^{2i \pi  \omega Q},
\ee
where $\lambda^{a'}$ are the hermitian generators and $Q$ is the predefined generator with a given direction in generator space, while $\omega$ and $\beta^{a'}$ are the corresponding parameters. Combining with the above consistency condition
and expanding infinitesimally one immediately finds the
condition [6],
\be \label{eq:baz}
 \left \{ \vec{\beta}\cdot\vec{\lambda},Z_2 \right\}=0 \quad  \textrm{and} \quad [T_g, Z_2]=0.
\ee
These relations determine the broken and unbroken parts of generators under consideration. The latter also gives the singular solution $T_g = \pm 1$.

For 5D theory compactified on the $S^1/Z_2$ orbifold with the Scherk-Schwarz twist in Eq. \eqref{eq:gsym}, the twisted field obeys,
\be \label{eq:tf}
    \phi(x,y+2 \pi R) = e^{2i \pi \omega Q} \phi(x,y),
\ee
where $R$ is the compactification radius. Symmetry breaking
is achieved if the symmetry generated by $Q$ is broken
by the 5D kinetic term and satisfies the anticommutative
relation while the unbroken parts generated by $Q'$ are determined by the second relation in Eq. \eqref{eq:baz} [6], that is
\be \label{eq:omqz}
   \left \{ \omega Q,Z_2 \right\}= \omega \left \{Q,Z_2 \right\}= 0 \quad \textrm{and} \quad [Q',Z_2] = 0.
\ee

\subsection{Orbifold breaking mechanisms in 5D \susix}
We are now ready to apply the preceding discussion on the orbifold $S^1/Z_2$ to \susix. $Z_2$ for \susix \, can be constructed based on 3 arrays of \sutwo \, type matrix along its diagonal elements as follows,
\be \label{zmtx}
Z_2=\begin{pmatrix}
    \begin{array}{cc}
       1 & 0 \\
       0 & 1
     \end{array}  &  &  \\
     & \begin{array}{cc}
       1 & 0 \\
       0 & -1
     \end{array} &  \\
     &  & \begin{array}{cc}
       -1 & 0 \\
       0 & -1
     \end{array} \\
  \end{pmatrix}.
\ee
This form satisfies the boundary conditions of orbifold $S^1/Z_2$ suitable to realize the symmetry breaking \susix$\rightarrow$\suthr$\otimes$\suthr$\otimes$\uone.

For $Z_2$ given in Eq. \eqref{zmtx}, the broken parts satisfying Eq. \eqref{eq:omqz} are the \susix \,  generators with off-diagonal elements, that is $\lambda^{\hat{a}}$ with $\hat{a} = 9,\cdots,26$. On the contrary $\lambda^{a}$, with $a = 1,\cdots,8, 27,\cdots,35$, determines the unbroken parts. [2]

Due to orbifold singular points, the parity operator $Z_2$ which operates at each singular point is labeled as $Z^{(0)}_{2}$ for $y=0$ and $Z^{(1)}_{2}$ for $y=\pi R$, $Z^{(0)}_{2}=Z^{(1)}_{2}$ as in Eq. \eqref{zmtx}, and the following relation $U = Z^{(0)}_{2}Z^{(1)}_{2}= I_{6}$ holds which also provides an alternative to Eq. \eqref{zmtx}, that is, $Z_2=U$.

Special conditions can easily be obtained from  Eq. \eqref{eq:omqz}  which determine two special breaking patterns of orbifold $S^1/Z_2$ i.e. trivial orbifold breaking and the pseudo non-trivial breaking. The first is dictated by commutator of  Eq. \eqref{eq:omqz} setting  $Z_2=U$, then $Q'$ comprises  all generators of \susix \, which are consequently unbroken leading  to dimensional deconstruction without gauge symmetry breaking [15]. The second is actually the special condition of a more general condition, with $\omega \neq 0$ and $Q \neq 0$, known as non-trivial orbifold breaking [15],  coming from anti-commutator of  Eq. \eqref{eq:omqz} where $Z_2$ as in Eq. \eqref{zmtx}, $\omega \neq 0$  and Q is set to zero. This provides no broken generator eventhough field is twisted which also leads to dimensional deconstruction with intact symmetry. Accordingly both conditions give the same breaking pattern 5D-\susix$\rightarrow$4D-\susix.

The next symmetry breaking is performed by so-called \susix \,  Little-like (Baby) Higgs which will be derived later with breaking pattern following the (pseudo) non-trivial one with $Z_2$ in Eq. \eqref{zmtx} giving 4D-\susix$\rightarrow$ 4D-\suthr$\times$\suthr$\times$\uone.

\subsection{The 5D-model and Lagrangian}
\subsubsection{The 5D-model with 2 branes}

\begin{figure}[h]
\centering
\includegraphics[width=5.5cm, height=3.5cm, keepaspectratio=true]{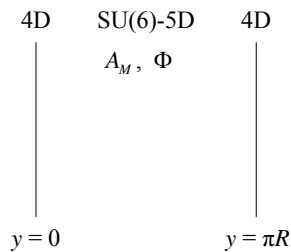}
\caption{\label{5D4Dbrane}\small{5D-model with 4D-particles live in 2-brane}.}
\end{figure}

We adopt the 5D-model with the 4D-particles live in 2 branes and 5D-gauge bosons as well as scalar bosons  live in the bulk. One brane corresponds to fixed point $y=0$ and the other brane corresponds to another fixed point $y= \pi R$ of the orbifold $S^1/Z_2$ as per Fig. \ref{5D4Dbrane}.

The boundary conditions consist of unitary operator $U$ and parity operator $Z_2$ which have the following  twisted boundary conditions
\begin{alignat}{2}\label{eq:AM}
     A_M(x,y+2\pi R)&= U A_M(x,y) U^{\dag}\\\label{eq:Amu}
      A_{\mu}(x,-y)&= Z_2 A_{\mu}(x,y)Z^{\dag}_2, \,\, A_y(x,-y)=-Z_2A_y(x,y)Z^{\dag}_2
\end{alignat}
while for fermion $\psi^{f}(x,-y)=\gamma_5Z_2\psi^{f}(x,y)$  where $U$ is unitary matrix free from $y, U \in T_g$ and $Z_2$ is $Z_2$-parity transformation matrix $(Z^2_2=I_N)$. Two sets of boundary conditions such as $(U, Z_0)$ and $(U', Z'_0)$ can be equivalent one to another in the sense that its physical content is the same \textit{i.e} $(U, Z_0)\equiv (U', Z'_0)$ provided the following conditions $\partial_MU'=0, \,  \partial_M Z'_0 =0, \, Z'^{\dag}_{0}=Z'_0$ are set [15].

Therefore Eq.\eqref{eq:AM} modifies itself as
\be \label{eq:AMxy2}
    A_M(x,y+2 \pi R) = A_M(x,y)
\ee

The 5D scalar boson can be obtained from Eq. \eqref{eq:tf}, setting $Q=0$ to provide Eq. \eqref{eq:tpxy2} below under the above pseudo non-trivial breaking,
\be \label{eq:tpxy2}
    \widetilde{\Phi}(x,y+2 \pi R) = \widetilde{\Phi}(x,y),
\ee
where $A_M$ is 5D \susix \, gauge bosons and $\widetilde{\Phi}$ is a single-valued and periodic function of 5D \susix \, scalar boson. From the general property of any field one can always obtain even and odd fields as follows,
\be \label{eq:Am}
    A_M=\tfrac{1}{2}(A_M+A'_M)+\tfrac{1}{2}(A_M-A'_M)=A_M^{(+)}+A_M^{(-)},
\ee
where the even: $ A_M^{(+)}(x,y)$ and $\widetilde{\Phi}_{+}(x,y)$ are separated from the odds, so-called parity splitting, under the requirement of pseudo non-trivial breaking.

The scalar boson in Eq.\eqref{eq:tpxy2} is the source of \susix-origin Little (Baby) Higgs boson which will produce \susix \, Baby Higgs that break  the 4D-\susix \, into the above symmetry and \susix \,  will-be-SimplestLittleHiggs scalar. Nevertheless one encounters the problem where Little Higgs is normally strongly-coupled near the cut-off scale of the theory while \susix \, GUT is weakly-coupled. On the other hand, the duality of orbifold breaking pattern, the trivial and the pseudo non-trivial, poses another problem due to opposite consequences of the breaking pattern. The trivial one demands for no scalar boson to exist in 4D \susix, while the pseudo non-trivial allows scalar boson. This seemingly contradictory condition is actually what is exactly needed.

The scalar boson after the compactifying with $y \sim 0$ provides the strongly-coupled scalar which lives near the \susix \, cut-off scale so-called \susix-origin Little (Baby) Higgs $\widetilde{\Phi}^{(i)}_{+}, i=1,2$. Under the requirement of trivial condition this cannot exist in 4D \susix \, GUT which is automatically accomplished due to decoupling of this \susix-origin Little(Baby) Higgs from the much lower-energy \susix \, GUT. This fact can also be understood from coupling constant, that is, strongly-coupled scalars shall not mix with weakly-coupled \susix \, GUT. In this way the non-existant scalar in trivial breaking is accomplished. On the other hand the demand of pseudo non-trivial breaking for scalar boson can be accomplished by adjusting \susix-origin Little(Baby) Higgs which consist of zero mode and higher modes to the 4D \susix \, GUT requirement. If higher modes are eliminated due to the fact that 4D-content mostly resides in the zero mode term then \susix \, Baby Higgs $\widetilde{\Phi}^{(i)}_{+,\textrm{P}}, i=1,2$  (weakly-coupled) is obtained by means of selecting the lowest-order expansion. Another possibility is expanding $e^{\left(\begin{smallmatrix}
                                                                   a & b \\
                                                                   c & d \\
                                                                 \end{smallmatrix}\right)} \sim \left(\begin{smallmatrix}
                                                                   e^a & e^b \\
                                                                   e^c & e^d \\
                                                                 \end{smallmatrix}\right)$
to result in \susix \, will-be-SimplestLittleHiggs scalar. In this way both requirement of trivial and pseudo non-trivial breaking are fulfilled while \susix \, Baby Higgs is also reconciled with \susix \, GUT quite well.

\subsubsection{The Origin of \susix \, scalar boson}
We can consider the 5D \susix \, scalar boson as a periodic function with the even component [16],
\be
     \widetilde{\Phi}_{+}(x,y)= \frac{1}{\sqrt{\pi R}}
                           \Phi^{0}_{+}(x)
                           +\sqrt{\frac{1}{\pi R}} \sum^{\infty}_{n=2}\Phi^{n}_{+}(x) \cos \left(\frac{ny}{R} \right)
\ee
while the odd component is,
\be
    \widetilde{\Phi}_{-}(x,y)=\sqrt{\frac{1}{\pi R}} \sum^{\infty}_{n=1}\Phi^{n}_{-}(x) \sin \left(\frac{ny}{R} \right).
\ee
The 5D \susix \, periodic scalar bosons must have two sets of bi-parity due to two orbifold fixed points $y=0$ and $y=\pi R$ with combination of parity  $(+,\pm)$ due to $(Z^{(0)}_{2},Z^{(1)}_{2})$ and $(-,\pm)$ due to the same $(Z^{(0)}_{2},Z^{(1)}_{2})$ where the first parity belongs to $y=0$ brane.

Therefore one can write 5D-\susix \, periodic scalar bosons as the even scalar [20],
\\
\begin{alignat}{2}\label{eq:wp++xy}
  \widetilde{\Phi}_{+,+}(x,y)
    &= \frac{1}{\sqrt{\pi R}}
                           \Phi^{0}_{+,+}(x)
                           +\sqrt{\frac{1}{\pi R}} \sum^{\infty}_{n=2}\Phi^{n}_{+,+}(x) \cos \left(\frac{ny}{R} \right),\\ \label{eq:wp+-xy}
     \widetilde{\Phi}_{+,-}(x,y)
     &= \frac{1}{\sqrt{\pi R}}
                           \Phi^{0}_{+,-}(x)
                           +\sqrt{\frac{1}{\pi R}} \sum^{\infty}_{n=2}\Phi^{n}_{+,-}(x) \cos \left(\frac{(n+\tfrac{1}{2})y}{R} \right)
\end{alignat}
and the odd scalars
\begin{alignat}{2}\label{}
  \widetilde{\Phi}_{-,+}(x,y)
    &= \sqrt{\frac{1}{\pi R}} \sum^{\infty}_{n=1}\Phi^{n}_{-,+}(x) \sin \left(\frac{(n+\tfrac{1}{2})y}{R} \right),\\
     \widetilde{\Phi}_{-,-}(x,y)
     &= \sqrt{\frac{1}{\pi R}}\sum^{\infty}_{n=1}\Phi^{n}_{-,-}(x) \sin \left(\frac{ny}{R} \right).
\end{alignat}
Considering the 4D-terms (parts) in Eqs. \eqref{eq:wp++xy} and \eqref{eq:wp+-xy} which can be viewed as serial terms of exponential expression one can consider  \susix \, scalar boson  which lives in 4D spacetime under the cut-off scale of \susix \, theory, $ \Lambda^{4\textrm{D}}_{(6)}$, to be defined in general form as follows,

\be \label{eq:p1+p2+}
\widetilde{\Phi}^{(1)}_{+} =v e^{\frac{if_2}{f_1}\theta} \quad \textrm{and} \quad
\widetilde{\Phi}^{(2)}_{+} =v' e^{-\frac{if_1}{f_2}\theta},
\ee
where $v$ and $v'$ are \susix \, VE$V$s and $\theta$ PNB parameter which are defined later.

Lets define near-brane area as area with small $y$ and $\alpha = \omega y/R$ where $\alpha$ is also relatively small then a global gauge transformation $e^{i\alpha}$ which works in periodic scalar field is obtained and produce the shift symmetry for PNB, $\frac{f_2}{f_1}\theta \rightarrow \frac{f_2}{f_1}\theta + [\alpha]$ and $\frac{f_1}{f_2}\theta \rightarrow \frac{f_1}{f_2}\theta - [\alpha]$ with $[\alpha]: 6 \times 6$ matrix containing $\alpha$, which protects the masslessness of PNB (and the \susix \, scalar boson). The shift symmetry forbids all other terms except kinetic terms in Lagrangian otherwise it is broken  [19] and the global symmetry as well [6]. But \susix \, global symmetry [20] breaking can be triggered by orbifold breaking due to extra-dimensional property of $\alpha$ in the lower-near-brane [6,19].

The Lagrangian can be written accordingly as
\be \label{eq:lsu65}
    \mathcal{L}^{\textrm{\susix}}_{5} = D^{M}\Phi^{\dag}D_{M}\Phi, \quad M=(\mu,y).
\ee

Scalar field $\Phi$ is expressed as periodic scalar field $\widetilde{\Phi}$ via the following relationship [6] where $\Phi=(\Phi_1, \Phi_2, \Phi_3, \Phi_4, \Phi_5, \Phi_6)^T \equiv [\Phi_k], k=1,2,\cdots, 6$ is scalar boson in the fundamental representation of \susix,

\be
    \Phi(x,y)=e^{i \omega Q_{v} y/R}\, \widetilde{\Phi}(x,y)= e^{i Q_{v} \alpha}\,\widetilde{\Phi}(x,y),
\ee
which can be obtained as solution of Eq. \eqref{eq:tf} and $Q_{v}$ represents \susix \, broken generators at the direction of VE$V$s [6,20]. Defining $D^{\mu}(D_{\mu})$  as 4D-covariant derivative and $D^{y}(D_{y})$ as fifth-dimensional covariant derivative with $T^{a}=\lambda^a/2 (=T_a)$
\be \label{eq:cder}
\begin{split}
  D_{\mu} &=\partial_{\mu}-ig_5A^{a}_{\mu}T_a, \, \,
  D^{\mu} =\partial^{\mu}+ig_5A^{\mu}_{a}T^a \, \, \, \textrm{and} \, \, \, \\
  D_{y} &=\partial_{y}+ig_5A^{a}_{y}T_a, \, \,
  D^{y} =\partial^{y}-ig_5A^{y}_{a}T^a, \\
\end{split}
\ee
where $g_5$ is the 5D coupling constant one can separate the 4D-brane from the bulk Lagrangian
 \be
        \mathcal{L}^{\textrm{\susix}}_{5} =  \mathcal{L}^{\textrm{brane}}_{\mu}+ \mathcal{L}^{\textrm{near-brane}}_{(\theta y),y}+ \mathcal{L}^{\textrm{\susix}}_{y},
 \ee
where 4D near-brane is just in-between brane and bulk.

Thus the Lagrangians,  setting $Q_{v}=0$ for \susix \,  upper-near-brane (thus eliminating Lorentz invariant-violating term), after Scherk-Schwarz but prior to orbifold breaking, can be expressed as follows,
\begin{alignat}{2}\label{eq:lsu6m}
    \mathcal{L}^{\textrm{brane}}_{\mu}
    &=D^{\mu}\widetilde{\Phi}^{\dag}D_{\mu}\widetilde{\Phi},\\\label{eq:lsu6ye}
     \mathcal{L}^{\textrm{\susix}}_{y}
    &=D^{y}\widetilde{\Phi}^{\dag}D_{y}\widetilde{\Phi}+ig_5(A^{a}_{y}T_a\widetilde{\Phi}D^{y}\widetilde{\Phi}^{\dag}
    -A^{y}_{a}T^a\widetilde{\Phi}^{\dag}D_{y}\widetilde{\Phi}).
\end{alignat}
while for $\mathcal{L}^{\textrm{near-brane}}_{(\theta y),y}$ two cases happen and are determined by shift-symmetry-breaking parameter $(\theta \alpha)$ (or $\theta y$ due to $\alpha = \frac{\omega y}{R}$) as follows:
In the upper-near-brane where shift symmetry is intact and ($\theta y$)-term is negligible so that $\theta$ is small (and also for the reason which will be clear after Eq.(28) where gauge-scalar unification [15] is applied) one finds, based on Eq.\eqref{eq:p1+p2+}, as follows
\be \nonumber
(y \sim 0, (\theta y) \rightarrow 0), \, D^{\mu}\widetilde{\Phi}^{\dag}= D_{\mu}\widetilde{\Phi}=0 \rightarrow \,  \textrm{due to} \, \, \partial^{\mu}\theta \, \, (\textrm{or} \, \,  \partial_{\mu}\theta) \rightarrow 0,
\ee
while in the lower-near-brane where shift symmetry is broken  ($\theta y$)-term and also $\theta$ are significant one finds the following,
\be \nonumber
(y \sim 0, (\theta y) \rightarrow \mu \neq 0), \, D^{\mu}\widetilde{\Phi}^{\dag}\neq D_{\mu}\widetilde{\Phi}= \textrm{significant value of} \, \,   \partial^{\mu}\theta (\partial_{\mu}\theta).
\ee
Adjusting notation for both lower- and upper-near-brane, by means of $\mathcal{L}^{\textrm{near-brane}}_{(\theta y),y} = \mathcal{L}^{\textrm{near-brane}}_{\mu}$ and $\mathcal{L}^{\textrm{near-brane}}_{(\theta y),y} =\mathcal{L}^{\textrm{near-brane}}_{y}$, one finally obtains, for upper-near-brane:
\be \label{eq:lnby}
\mathcal{L}^{\textrm{near-brane}}_{y}= \delta(y) \left\{ \tfrac{1}{2}g^2_5\widetilde{\Phi}^{\dag}A^{y}_{a}A^{a}_{y}\widetilde{\Phi} \right\},
\ee
(and for lower-near-brane: for the reason which will be clear after Eq.(28) and Subsection 3.3 as,
\be \nonumber
\mathcal{L}^{\textrm{near-brane}}_{\mu} = V^{(6)}_{\mu},
\ee
where $ V^{(6)}_{\mu}$ will turn out to be $V^{(6)}_{\mu \textrm{P}}$ in Eq.(50)).

Eq. \eqref{eq:lsu6m} reflects the condition in the brane $(y=0, \mu)$  while Eq. \eqref{eq:lsu6ye} of the far-distant out-of-brane condition $(y>0, \mu)$.

\section{Near-Brane 4D Scalar Bosons}
\subsection{Upper-near-brane extra-strongly-coupled scalar boson (\susix-origin Little(Baby) Higgs scalar boson)}
In the brane $y=0$ and near-brane $y \sim 0$ the even scalar bosons in Eqs. \eqref{eq:wp++xy} and \eqref{eq:wp+-xy} become as
\begin{alignat}{2}\label{}
    \widetilde{\Phi}^{(1)}_{+}(x) &= \widetilde{\Phi}_{+,+}(x,y)\mid_{y=0 \, \textrm{or} \sim 0}, \quad
     \widetilde{\Phi}^{(2)}_{+}(x) &= \widetilde{\Phi}_{+,-}(x,y)\mid_{y=0 \, \textrm{or} \sim 0}.
\end{alignat}
For the upper-near-brane area Neumann boundary condition dictates the following $D^{y}\widetilde{\Phi}^{\dag}=D_{y}\widetilde{\Phi}=0$ and based on property of extra-dimensional dominance $D^{\mu}\widetilde{\Phi}^{\dag}=D_{\mu}\widetilde{\Phi}=0$  make Eqs. \eqref{eq:lsu6m} and \eqref{eq:lsu6ye} zero. This shows that Eq. \eqref{eq:lnby} is really the only upper-near-brane equation with $\delta(y)=1$ for $y \sim 0$ and under both trivial and pseudo non-trivial breaking condition, $Q_{v}=0$, the final value has been obtained,
\be \label{eq:lsu6y}
    \mathcal{L}^{\textrm{near-brane}}_{y}
    =\tfrac{1}{2}g^2_5\left(\widetilde{\Phi}^{(i)\dag}A^{y}_{a}\right)\left(A^{a}_{y}\widetilde{\Phi}^{(i)}\right),
\ee
where now $\widetilde{\Phi}^{(i)}=\widetilde{\Phi}^{(i)}_{+}(x)$ and $\widetilde{\Phi}^{(i)\dag}=\widetilde{\Phi}^{(i)\dag}_{+}(x)$, with $i=1,2$.

In this upper-near-brane  bulk $(y-\textrm{area})$, under the provision of pseudo non-trivial orbifold breaking where $A^{y}_{\hat{a}} T^{\hat{a}}$  and $A_{y}^{\hat{a}} T_{\hat{a}}$ produce upper-near-brane scalar due to gauge-scalar unification [15], one has the subsets  (sextet out of $2 \times 9$ broken $A^{y}_{\hat{a}}$ and $A_{y}^{\hat{a}}$) from Eq. \eqref{eq:Amu}
\be
    A^{y}_{\hat{a}}T^{\hat{a}} \supset \widetilde{\Phi}^{(j)}, \quad
    A_{y}^{\hat{a}}T_{\hat{a}} \supset \widetilde{\Phi}^{(j)\dag}
\ee
where $\widetilde{\Phi}^{(j)} (\textrm{or} \, \widetilde{\Phi}^{(j)\dag})$  is diagonal $3\times3$ sub-matrix component of $6\times6$ matrix of $A^{y}_{\hat{a}}T^{\hat{a}}\, (\textrm{or} \, A_{y}^{\hat{a}}T_{\hat{a}}) $ and $j=1,2$ due to hermitian conjugacy and the following
$D^{\mu}\widetilde{\Phi}^{\dag}=D^{\mu}A_{y}^{\hat{a}}T_{\hat{a}}=0$ and $D_{\mu}\widetilde{\Phi}=D_{\mu}A^{y}_{\hat{a}}T^{\hat{a}}=0$. On the other side in the lower-near-brane $D^{\mu}\widetilde{\Phi}^{\dag} \neq D_{\mu}\widetilde{\Phi}\neq 0$ due to significant $\partial^{\mu}\theta (\partial_{\mu}\theta)$ because of dominant 4D-property. The only  term of Eq. \eqref{eq:lsu6y} provides the quartic term with index $(i)$ to label the original scalar boson in Eq. \eqref{eq:lsu6y} which can be rewritten, using $\widetilde{\Phi}^{(j)}$ as diagonal component of $A^{y}_{\hat{a}}T^{\hat{a}}$, as
\be \label{eq:lterm}
    V^{(6)}_{y}=\lambda^{(6)}_{y}(\widetilde{\Phi}^{(i)\dag}\widetilde{\Phi}^{(j)})
    (\widetilde{\Phi}^{(j)\dag}\widetilde{\Phi}^{(i)})
\ee
with $\lambda^{(6)}_{y}= g^2_5$. If one takes $g_5 \sim \mathcal{O}(1)$ then $\lambda^{(6)}_{y}\sim \mathcal{O}(1)$ to $\mathcal{O}(10)$ in Eq.\eqref{eq:lterm} which reflects a very strong quartic coupling of typical Coleman-Weinberg potential. Since $\widetilde{\Phi}^{(j)}(\widetilde{\Phi}^{(j)\dag})$ is arbitrarily  taken from $9$ broken $A^{y}_{\hat{a}}T^{\hat{a}}$ (and $9$ broken $A_{y}^{\hat{a}}T_{\hat{a}}$) it is justified to take $i\neq j$ in Eq. \eqref{eq:lterm}.

In the near-brane with extra-strongly-interacting  \susix \, scalars,  the potential in Eq. \eqref{eq:lterm}  is zero  as required by shift symmetry on Pseudo Nambu-Goldstone Boson (PNB), $\theta$. Therefore one can call $\widetilde{\Phi}_{+}^{(1)}(x)$ and $\widetilde{\Phi}_{+}^{(2)}(x)$ as the \susix-origin Little (Baby)  Higgs which must be defined by, making use the most basic original parameters, radius of compactification $R$,  two Higgs doublets $h$ and $h'$ and two Scherk-Schwarz parameters $\omega_1$ and $\omega_2$ following $i\neq j=1,2$. Here can one also defines two VE$V$s, $v$ and $v'$ in accordance to  two Scherk-Schwarz parameters at near-brane $y \sim 0$  as,
\be \label{eq:vf1}
    v= \left(\begin{matrix}
    0 \\
    0 \\
    \frac{\omega_1}{R} \\
    0 \\
    0 \\
    0 \\
    \end{matrix}\right)= \frac{1}{\sqrt{\pi R}} \left(\begin{matrix}
    0 \\
    0 \\
    f_1 \\
    0 \\
    0 \\
    0 \\
    \end{matrix}\right),  v'= \left(\begin{matrix}
    0 \\
    0 \\
    0 \\
    0 \\
    0 \\
    \frac{\omega_2}{R} \\
    \end{matrix}\right)= \frac{1}{\sqrt{\pi R}} \left(\begin{matrix}
    0 \\
    0 \\
    0 \\
    0 \\
    0 \\
    f_2 \\
    \end{matrix}\right),
\ee
\be \label{eq:vf2}
   f_1=\frac{\omega_1 \sqrt{\pi}}{\sqrt{R}}, f_2=\frac{\omega_2 \sqrt{\pi}}{\sqrt{R}}.
\ee

The parameterization of \susix-origin  Little (Baby) Higgs is governed
by the number of scalar doublets which are allowed to
be put in $6 \times 6$ matrix. Thus it depends on the number
of generated PNBs through the condition $a'_{jk}\widetilde{\Phi}_{k} \neq 0$ with $a' = 1,\cdots, 35$ and $\langle\widetilde{\Phi}^{(1)}\rangle = v$,  $\langle\widetilde{\Phi}^{(2)}\rangle = v'$. These determine the total number of PNBs to be $22$. However, the simplest Little Higgs at \suthr$\times$\suthr \,  requires only $10$ PNBs while \sutwo$\times$\uone \, produces $4$ PNBs to become \uone. Therefore one may yet have free 8 scalar bosons which could create $4$ scalar doublets to be assigned as the \susix-origin  Little (Baby) Higgs as follows,
\be \label{theta1/f}
    \theta= \frac{1}{f}
                         \left(\begin{smallmatrix}
                           (0)_{3 \times 3} & \left( \begin{smallmatrix}
                                               \begin{smallmatrix}
                                                   (0)_{2 \times 2} & (h)_{2 \times 1} \\
                                                   (h'^{\dag})_{1 \times 2} & 0 \\
                                                \end{smallmatrix}
                                                        \end{smallmatrix} \right) \\
                          \left(\begin{smallmatrix}
                                               \begin{smallmatrix}
                                                  (0)_{2 \times 2} & (h')_{2 \times 1} \\
                                                   (h^{\dag})_{1 \times 2} & 0 \\
                                                \end{smallmatrix}
                                                    \end{smallmatrix}\right)  & (0)_{3 \times 3} \\
                         \end{smallmatrix}\right)
\ee
where $f^2 = f^2_1 + f^2_2$. The scalar doublets $h$ and $h'$ are the would-be SM Higgs as will be clarified later.

Therefore Eq. \eqref{eq:p1+p2+} can be reexpressed in complete forms, for $\widetilde{\Phi}^{(1)}_{+} $ and $\widetilde{\Phi}^{(2)}_{+}$ successively, as

\be \label{eq:P1+P2+s}
      \frac{1}{\sqrt{\pi R}}e^{\frac{if_2}{f_1 f}  \left(\begin{smallmatrix}
                           (0)_{3 \times 3} & \left( \begin{smallmatrix}
                                               \begin{smallmatrix}
                                                   0 & 0 \\
                                                   0 & 0 \\
                                                \end{smallmatrix}
                                                          & h \\
                                                 h'^{\dag} & 0 \\
                                               \end{smallmatrix} \right)
 \\
                           \left(\begin{smallmatrix}
                                               \begin{smallmatrix}
                                                   0 & 0 \\
                                                   0 & 0 \\
                                                \end{smallmatrix}
                                                          & h' \\
                                                 h^{\dag} & 0 \\
                                               \end{smallmatrix}\right)  & (0)_{3 \times 3} \\
                         \end{smallmatrix}\right)}
 \left(\begin{matrix}
    0 \\
    0 \\
    f_1 \\
    0 \\
    0 \\
    0 \\
    \end{matrix}\right) \, \,
\textrm{and} \, \,  \frac{1}{\sqrt{\pi R}}e^{-\frac{if_1}{f_2 f}  \left(\begin{smallmatrix}
                           (0)_{3 \times 3} & \left( \begin{smallmatrix}
                                               \begin{smallmatrix}
                                                   0 & 0 \\
                                                   0 & 0 \\
                                                \end{smallmatrix}
                                                          & h \\
                                                 h'^{\dag} & 0 \\
                                               \end{smallmatrix} \right)
 \\
                           \left(\begin{smallmatrix}
                                               \begin{smallmatrix}
                                                   0 & 0 \\
                                                   0 & 0 \\
                                                \end{smallmatrix}
                                                          & h' \\
                                                 h^{\dag} & 0 \\
                                               \end{smallmatrix}\right)  & (0)_{3 \times 3} \\
                         \end{smallmatrix}\right)}
 \left(\begin{matrix}
    0 \\
    0 \\
    0 \\
    0 \\
    0 \\
    f_2 \\
    \end{matrix}\right).
\ee
Under the requirement of shift symmetry \susix-origin  Little (Baby) Higgs scalar remains massless due to masslessness of PNB. The breaking of shift symmetry leads to the massiveness of PNB.

From Eq.\eqref{eq:p1+p2+} and Eq. \eqref{eq:P1+P2+s} it is clear $\widetilde{\Phi}^{(i)\dag}_{+}\widetilde{\Phi}^{(j)}=0, i \neq j=1,2 $ and quartic potential in Eq.\eqref{eq:lterm} becomes zero. This shows the \susix-origin Little (Baby) Higgses are still massless and interact one to another by means of quantum interaction and is called Extra-Strongly-Coupled (ESC). Here minimum potential is always zero so that the introduction of VE$V$s does not bring about \susix \, nor shift symmetry breaking. Reviewing again the higher terms of $e^{i\left(\frac{f_2}{f_1}\theta + \alpha Q_{v}\right)}$ or $e^{-i\left(\frac{f_1}{f_2}\theta - \alpha Q_{v} \right)}$ and making use Eq. \eqref{theta1/f} one finds mixed terms-$\alpha\theta Q_{v} $ (with factor $\frac{\alpha}{f}$) while neglecting $\alpha^{2}Q^{2}_{v}, \theta^{2}, (\theta+\alpha Q_{v})^3$ etc, which gives two interval of values i.e $\frac{\alpha}{f}$ neglectable or $\frac{\alpha}{f}$ significant. The first happens in upper (farther) part of near-brane while the second shows up in the lower (nearer) part of near-brane i.e shift symmetry is broken in the lower-near-brane.

If we expand $\widetilde{\Phi}^{(1)}_{+}$ and $\widetilde{\Phi}^{(2)}_{+}$ above the expressions in Eq.\eqref{eq:wp++xy} and Eq. \eqref{eq:wp+-xy}  are obtained immediately for $y=0 (y \sim 0)$ with zero mode and higher modes defined as follows,
\\
\begin{alignat}{3}\label{}
\textrm{for}\, \,  \tilde{\Phi}^{1}_{+}:  \tilde{\Phi}^{0}_{+,+}(x)&=
                         \left\{1 + \frac{if_2}{f_1f}
                         \left(\begin{smallmatrix}
                           (0)_{3 \times 3} & \left( \begin{smallmatrix}
                                               \begin{smallmatrix}
                                                   0 & 0 \\
                                                   0 & 0 \\
                                                \end{smallmatrix}
                                                          & h \\
                                                 h'^{\dag} & 0 \\
                                               \end{smallmatrix} \right)
 \\
                           \left(\begin{smallmatrix}
                                               \begin{smallmatrix}
                                                   0 & 0 \\
                                                   0 & 0 \\
                                                \end{smallmatrix}
                                                          & h' \\
                                                 h^{\dag} & 0 \\
                                               \end{smallmatrix}\right)  & (0)_{3 \times 3} \\
                         \end{smallmatrix}\right)\right\}
         \left(\begin{smallmatrix}
                                        0 \\
                                        0 \\
                                        f_1 \\
                                        0 \\
                                        0 \\
                                        0 \\
                                      \end{smallmatrix}\right),\\
 \textrm{for}\, \,  \tilde{\Phi}^{2}_{+}:  \widetilde{\Phi}^{0}_{+,-}(x)&=
                         \left\{1 - \frac{if_1}{f_2f}
                         \left(\begin{smallmatrix}
                           (0)_{3 \times 3} & \left( \begin{smallmatrix}
                                               \begin{smallmatrix}
                                                   0 & 0 \\
                                                   0 & 0 \\
                                                \end{smallmatrix}
                                                          & h \\
                                                 h'^{\dag} & 0 \\
                                               \end{smallmatrix} \right)
 \\
                           \left(\begin{smallmatrix}
                                               \begin{smallmatrix}
                                                   0 & 0 \\
                                                   0 & 0 \\
                                                \end{smallmatrix}
                                                          & h' \\
                                                 h^{\dag} & 0 \\
                                               \end{smallmatrix}\right)  & (0)_{3 \times 3} \\
                         \end{smallmatrix}\right)\right\}
         \left(\begin{smallmatrix}
                                        0 \\
                                        0 \\
                                        0 \\
                                        0 \\
                                        0 \\
                                        f_2 \\
                                      \end{smallmatrix}\right),\\ \nonumber
 \textrm{and for both:}& \\
 \widetilde{\Phi}^{n}_{+,+}(x)\left[\widetilde{\Phi}^{n}_{+,-}(x) \right]&=\frac{1}{n!}
                         \left\{\frac{if_2}{f_1f}\left[-\frac{if_1}{f_2f} \right]
                         \left(\begin{smallmatrix}
                           (0)_{3 \times 3} & \left( \begin{smallmatrix}
                                               \begin{smallmatrix}
                                                   0 & 0 \\
                                                   0 & 0 \\
                                                \end{smallmatrix}
                                                          & h \\
                                                 h'^{\dag} & 0 \\
                                               \end{smallmatrix} \right)
 \\
                           \left(\begin{smallmatrix}
                                               \begin{smallmatrix}
                                                   0 & 0 \\
                                                   0 & 0 \\
                                                \end{smallmatrix}
                                                          & h' \\
                                                 h^{\dag} & 0 \\
                                               \end{smallmatrix}\right)  & (0)_{3 \times 3} \\
                         \end{smallmatrix}\right)\right\}^n
         \left(\begin{smallmatrix}
                                        0 \\
                                        0 \\
                                        f_1 \\
                                        0 \\
                                        0 \\
                                        0 \\
                                      \end{smallmatrix}\right)\left[\left(\begin{smallmatrix}
                                        0 \\
                                        0 \\
                                        0 \\
                                        0 \\
                                        0 \\
                                        f_2 \\
                                      \end{smallmatrix}\right)\right], \\ \nonumber  & \quad n=2,3,\dots,\infty.
\end{alignat}

These give perturbative expressions of \susix-origin Little (Baby) Higgses from which the so-called \susix \, Baby Higgses are defined in this paper as lowest-order of the expansion i.e. the zero mode. Consequently one must establish a cut-off scale for perturbative approach $\Lambda^{\textrm{NP}}_{(6)}$ above which only the ESC \susix-origin Little(Baby) Higgs theory takes control i.e. $\Lambda^{\textrm{NP}}_{(6)}< \textrm{ESC regime (upper part)} < \Lambda^{\textrm{4D}}_{(6)}$.

\subsection{Lower-near-brane strongly-coupled scalar boson (\susix \, will-be-SimplestLittleHiggs scalar boson)}
Now lets discuss the mechanisms to generate Simplest Little Higgs from \susix-origin Little(Baby) Higgs scalar in Eq. \eqref{eq:P1+P2+s}, assuming $\mathcal{O}(f) \alpha \mathcal{O}(f_{i})$ yields $f_2/f_1f$ and $f_1/f_2f \ll 1$. Then can one perform the following expansion
\be \label{eq:gfsu61}
      e^{\frac{if_2}{f_1 f}\left[-\frac{if_1}{f_2 f} \right] \left(\begin{smallmatrix}
                           (0)_{3 \times 3} & \left( \begin{smallmatrix}
                                               \begin{smallmatrix}
                                                   0 & 0 \\
                                                   0 & 0 \\
                                                \end{smallmatrix}
                                                          & h \\
                                                 h'^{\dag} & 0 \\
                                               \end{smallmatrix} \right)
 \\
                           \left(\begin{smallmatrix}
                                               \begin{smallmatrix}
                                                   0 & 0 \\
                                                   0 & 0 \\
                                                \end{smallmatrix}
                                                          & h' \\
                                                 h^{\dag} & 0 \\
                                               \end{smallmatrix}\right)  & (0)_{3 \times 3} \\
                         \end{smallmatrix}\right)} \sim
 \begin{pmatrix}
   1 & e^{\frac{if_2}{f_1 f}\left[-\frac{if_1}{f_2 f} \right]  \left(\begin{smallmatrix}
                                               \begin{smallmatrix}
                                                   0 & 0 \\
                                                   0 & 0 \\
                                                \end{smallmatrix}
                                                          & h \\
                                                 h'^{\dag} & 0 \\
                                               \end{smallmatrix}\right)} \\
   e^{\frac{if_2}{f_1 f}\left[-\frac{if_1}{f_2 f} \right]  \left(\begin{smallmatrix}
                                               \begin{smallmatrix}
                                                   0 & 0 \\
                                                   0 & 0 \\
                                                \end{smallmatrix}
                                                          & h' \\
                                                 h^{\dag} & 0 \\
                                               \end{smallmatrix}\right)} & 1 \\
 \end{pmatrix}
\ee
to obtain as follows,
\begin{alignat}{2}\label{eq:phi1+ph2}
    \widetilde{\Phi}^{(1)'}_{+}&=\begin{pmatrix}
                    \phi^{(1)}_{0} \\
                    \phi^{(1)} \\
                  \end{pmatrix}  \quad \textrm{and} \quad
    \widetilde{\Phi}^{(2)'}_{+}&=\begin{pmatrix}
                     \phi^{(2)} \\
                    \phi^{(2)}_{0} \\
                  \end{pmatrix},
\end{alignat}
where $\widetilde{\Phi}^{(i)'}_{+}, i=1,2$ is the \susix \, will-be-SimplestLittleHiggs scalar and lives below the scale $\Lambda^{\textrm{NP}}_{(6)}$,
\begin{alignat}{2}\label{eq:p10}
    \phi^{(1)}_{0}&=\frac{1}{\sqrt{\pi R}}\begin{pmatrix}
                                           0 \\
                                           0 \\
                                           f_1 \\
                                         \end{pmatrix}, \quad
    \phi^{(1)}&=\frac{1}{\sqrt{\pi R}}e^{\frac{if_2}{f_1f} \left(\begin{smallmatrix}
                                               \begin{smallmatrix}
                                                   0 & 0 \\
                                                   0 & 0 \\
                                                \end{smallmatrix}
                                                          & h' \\
                                                 h^{\dag} & 0 \\
                                               \end{smallmatrix}\right)}\begin{pmatrix}
                                           0 \\
                                           0 \\
                                           f_1 \\
                                         \end{pmatrix},\\
    \phi^{(2)}_{0}&=\frac{1}{\sqrt{\pi R}}\begin{pmatrix}
                                           0 \\
                                           0 \\
                                           f_2 \\
                                         \end{pmatrix}, \quad
    \phi^{(2)}&=\frac{1}{\sqrt{\pi R}}e^{-\frac{if_1}{f_2f} \left(\begin{smallmatrix}
                                               \begin{smallmatrix}
                                                   0 & 0 \\
                                                   0 & 0 \\
                                                \end{smallmatrix}
                                                          & h \\
                                                 h'^{\dag} & 0 \\
                                               \end{smallmatrix}\right)}\begin{pmatrix}
                                           0 \\
                                           0 \\
                                           f_2 \\
                                         \end{pmatrix}.
\end{alignat}
This must  happen after $f$ reaches significantly lower value than $\Lambda^{\textrm{NP}}_{(6)}$ in the lower-near-brane so that $\frac{\alpha}{f}$ of $\alpha \theta$-term becomes significant and shift symmetry is broken. Massive pseudo Nambu-Goldstone boson (PNB) is absorbed by scalar to become VE$V$, $\phi^{(1)}_{0}$ and $\phi^{(2)}_{0}$. Defining the scale $f'_i=\frac{1}{\sqrt{\pi R}}f, H(H')=\frac{1}{\sqrt{\pi R}} h(h')$ finally one can rewrite the \suthr \, Little Higgs as,
\be \label{p1np2}
\phi^{(1)}=e^{\frac{if'_2}{f'_1f'} \left(\begin{smallmatrix}
                                               \begin{smallmatrix}
                                                   0 & 0 \\
                                                   0 & 0 \\
                                                \end{smallmatrix}
                                                          & H' \\
                                                 H^{\dag} & 0 \\
                                               \end{smallmatrix}\right)}\begin{pmatrix}
                                           0 \\
                                           0 \\
                                           f'_1 \\
                                         \end{pmatrix}, \quad
\phi^{(2)}=e^{-\frac{if'_1}{f'_2f'} \left(\begin{smallmatrix}
                                               \begin{smallmatrix}
                                                   0 & 0 \\
                                                   0 & 0 \\
                                                \end{smallmatrix}
                                                          & H \\
                                                 H'^{\dag} & 0 \\
                                               \end{smallmatrix}\right)}\begin{pmatrix}
                                           0 \\
                                           0 \\
                                           f'_2 \\
                                         \end{pmatrix}.
\ee
In case of $H=H'$, Eq. \eqref{p1np2} become basically the Simplest Little Higgs on \suthr$\times$\suthr \,  as expected [8].

\subsection{Lower-near-brane  Weakly-coupled scalar boson (\susix \,  Baby Higgs)}
This is represented by \susix \, Baby Higgses which are defined by zero mode approximation where perturbative approach has been taken up to lowest order. This scalar lives below energy scale $\Lambda^{\textrm{NP}}_{(6)}$. \susix \, Baby Higgses can be written as (P : perturbative),
\be \label{eq:p1p+p2+p}
\widetilde{\Phi}^{(1)}_{+,\textrm{P}}(x) =v \left(1+\frac{if_2}{f_1}\theta(x)\right), \quad
\widetilde{\Phi}^{(2)}_{+,\textrm{P}}(x) =v' \left(1-\frac{if_1}{f_2}\theta(x)\right),
\ee
or, in complete forms as follows
\begin{alignat}{2}\label{eq:wp1+Px}
 \widetilde{\Phi}^{(1)}_{+,\textrm{P}}(x)&=\frac{1}{\sqrt{\pi R}}
                         \left\{1 + \frac{if_2}{f_1f}
                         \left(\begin{smallmatrix}
                           (0)_{3 \times 3} & \left( \begin{smallmatrix}
                                               \begin{smallmatrix}
                                                   0 & 0 \\
                                                   0 & 0 \\
                                                \end{smallmatrix}
                                                          & h \\
                                                 h'^{\dag} & 0 \\
                                               \end{smallmatrix} \right)
 \\
                           \left(\begin{smallmatrix}
                                               \begin{smallmatrix}
                                                   0 & 0 \\
                                                   0 & 0 \\
                                                \end{smallmatrix}
                                                          & h' \\
                                                 h^{\dag} & 0 \\
                                               \end{smallmatrix}\right)  & (0)_{3 \times 3} \\
                         \end{smallmatrix}\right)\right\}
         \left(\begin{smallmatrix}
                                        0 \\
                                        0 \\
                                        f_1 \\
                                        0 \\
                                        0 \\
                                        0 \\
                                      \end{smallmatrix}\right),\\ \label{eq:wp2+Px}
\widetilde{\Phi}^{(2)}_{+,\textrm{P}}(x)&=\frac{1}{\sqrt{\pi R}}
                         \left\{1 - \frac{if_1}{f_2f}
                         \left(\begin{smallmatrix}
                           (0)_{3 \times 3} & \left( \begin{smallmatrix}
                                               \begin{smallmatrix}
                                                   0 & 0 \\
                                                   0 & 0 \\
                                                \end{smallmatrix}
                                                          & h \\
                                                 h'^{\dag} & 0 \\
                                               \end{smallmatrix} \right)
 \\
                           \left(\begin{smallmatrix}
                                               \begin{smallmatrix}
                                                   0 & 0 \\
                                                   0 & 0 \\
                                                \end{smallmatrix}
                                                          & h' \\
                                                 h^{\dag} & 0 \\
                                               \end{smallmatrix}\right)  & (0)_{3 \times 3} \\
                         \end{smallmatrix}\right)\right\}
         \left(\begin{smallmatrix}
                                        0 \\
                                        0 \\
                                        0 \\
                                        0 \\
                                        0 \\
                                        f_2 \\
                                      \end{smallmatrix}\right),
\end{alignat}
or, making use VE$V$s in Eq. \eqref{eq:vf1},
\begin{alignat}{2}\label{eq:p1+px}
    \widetilde{\Phi}^{(1)}_{+,\textrm{P}}(x)&=v+\frac{1}{\sqrt{\pi R}}
    \begin{pmatrix}
      0 \\
      0 \\
      0 \\
      \frac{if_2}{f_1f} \left(\begin{smallmatrix}
                                               \begin{smallmatrix}
                                                   0 & 0 \\
                                                   0 & 0 \\
                                                \end{smallmatrix}
                                                          & h' \\
                                                 h^{\dag} & 0 \\
                                               \end{smallmatrix}\right) \left(\begin{smallmatrix}
                                        0 \\
                                        0 \\
                                        f_1 \\
                                      \end{smallmatrix}\right) \\
    \end{pmatrix},
    \widetilde{\Phi}^{(2)}_{+,\textrm{P}}(x)&=v'-\frac{1}{\sqrt{\pi R}}
    \begin{pmatrix}
       \frac{if_1}{f_2f} \left(\begin{smallmatrix}
                                               \begin{smallmatrix}
                                                   0 & 0 \\
                                                   0 & 0 \\
                                                \end{smallmatrix}
                                                          & h \\
                                                 h'^{\dag} & 0 \\
                                               \end{smallmatrix}\right) \left(\begin{smallmatrix}
                                        0 \\
                                        0 \\
                                        f_2 \\
                                      \end{smallmatrix}\right) \\
                                      0 \\
      0 \\
      0 \\
    \end{pmatrix}.
\end{alignat}
Eq. \eqref{eq:p1+px}  brings us immediately to the orbifold-based field redefinition as follows,
\begin{alignat}{2}\label{eq:p12'+P}
     \widetilde{\Phi}^{(1)'}_{+,\textrm{P}}(x)&=\widetilde{\Phi}^{(1)}_{+,\textrm{P}}(x)-v+v', \quad
     \widetilde{\Phi}^{(2)'}_{+,\textrm{P}}(x)&=\widetilde{\Phi}^{(2)}_{+,\textrm{P}}(x)-v'+v.
\end{alignat}
The new \susix \,  Baby Higgses are surprisingly split into triplets of \suthr \, Little-like Higgses in accordance to ($x$ is not written for simplicity),
\begin{alignat}{2}\label{eq:phi1a+p}
    \widetilde{\Phi}^{(1)'}_{+,\textrm{P}}(x)&=\begin{pmatrix}
                    0_{3 \times 1} \\
                    \phi^{(1)}_{P} \\
                  \end{pmatrix}, \quad
    \widetilde{\Phi}^{(2)'}_{+,\textrm{P}}(x)&=\begin{pmatrix}
                     \phi^{(2)}_{P} \\
                    0_{3 \times 1} \\
                  \end{pmatrix},
\end{alignat}
where \suthr \,  Little-like Higgses triplets are defined and obtained as
\begin{alignat}{2}\label{eq:p1p}
    \phi^{(1)}_{\textrm{P}}&=\frac{1}{\sqrt{\pi R}}\left[
          \left\{\left(1+\frac{\Delta f}{f_1} \right)+\frac{if_2}{f_1f} \left(\begin{smallmatrix}
                                               \begin{smallmatrix}
                                                   0 & 0 \\
                                                   0 & 0 \\
                                                \end{smallmatrix}
                                                          & h' \\
                                                 h^{\dag} & 0 \\
                                               \end{smallmatrix}\right) \right\} \left(\begin{smallmatrix}
                                        0 \\
                                        0 \\
                                        f_1 \\
                                      \end{smallmatrix}\right) \right], \\ \label{eq:p2p}
    \phi^{(2)}_{\textrm{P}}&=\frac{1}{\sqrt{\pi R}}
    \left[\left\{\left(1-\frac{\Delta f}{f_2} \right)-
       \frac{if_1}{f_2f} \left(\begin{smallmatrix}
                                               \begin{smallmatrix}
                                                   0 & 0 \\
                                                   0 & 0 \\
                                                \end{smallmatrix}
                                                          & h \\
                                                 h'^{\dag} & 0 \\
                                               \end{smallmatrix}\right) \right\} \left(\begin{smallmatrix}
                                        0 \\
                                        0 \\
                                        f_2 \\
                                      \end{smallmatrix}\right) \right],
\end{alignat}
where $\Delta f=f_2-f_1$.

Eq. \eqref{eq:p1p+p2+p} shows, under global gauge transformation $e^{i\alpha Q_{v}}$ ($\alpha$ small), the conservation of shift symmetry due to negligible $\alpha Q_{v} \theta$, that is $\frac{f_2}{f_1} \alpha Q_{v} \theta \ll 1, \frac{f_1}{f_2} \alpha Q_{v} \theta \ll 1$ in the following terms $\left\{\left(1+\frac{if_2}{f_1}\theta \right)+i\alpha Q_{v} -\frac{f_2}{f_1} \alpha Q_{v} \theta\right\}$ and $\left\{\left(1-\frac{if_1}{f_2}\theta \right)+i\alpha Q_{v} +\frac{f_1}{f_2} \alpha Q_{v} \theta\right\}$. \susix \, Baby Higgs remains massless until field redefinition is performed which lowers down \susix \,  to \suthr \, scale. Shift symmetry is broken via Eqs. \eqref{eq:phi1a+p}-\eqref{eq:p2p} which make $\alpha Q_{v} \theta$-term becoming significant with respect to $\left( f_2/f_1\theta+\alpha Q_{v} \right)$ and $\left( f_1/f_2\theta-\alpha Q_{v}\right)$, so PNB  gets mass when global symmetry is broken in the lower-near-brane.

The potential of \susix \,  Baby Higgses follows from Eq. \eqref{eq:lterm} by replacing $\lambda^{(6)}_{y}\rightarrow \lambda^{(6)}_{\mu\textrm{P}}$,
\be \label{eq:v6yp}
    V^{(6)}_{\mu\textrm{P}}=\delta_{ij}\lambda^{(6)}_{\mu\textrm{P}}\widetilde{\Phi}^{(i)'\dag}_{+,\textrm{P}}
                          \widetilde{\Phi}^{(j)'}_{+,\textrm{P}}
                          \widetilde{\Phi}^{(j)'\dag}_{+,\textrm{P}}
                          \widetilde{\Phi}^{(i)'}_{+,\textrm{P}},
\ee
where it is clear from Eqs. \eqref{eq:phi1a+p}, \eqref{eq:p1p}  and \eqref{eq:p2p} that $\widetilde{\Phi}^{(1)'}_{+,\textrm{P}} \sim \left(\begin{smallmatrix}
       0_{3 \times 1} \\
       \left(\begin{smallmatrix}
          0 \\
          0 \\
          f_1/\sqrt{\pi R} \\
        \end{smallmatrix}\right)\\
     \end{smallmatrix} \right)$ and $\widetilde{\Phi}^{(2)'}_{+,\textrm{P}} \sim \left(\begin{smallmatrix}
     \left(\begin{smallmatrix}
          0 \\
          0 \\
          f_2/\sqrt{\pi R} \\
        \end{smallmatrix}\right)\\
       0_{3 \times 1} \\
            \end{smallmatrix} \right)$  for $i$ and $j=1,2$. One concludes that
$\widetilde{\Phi}^{(i)'}_{+,\textrm{P}}=\widetilde{\Phi}^{(j)'}_{+,\textrm{P}}$ for $i=j$ and $\widetilde{\Phi}^{(i)'}_{+,\textrm{P}} \sim \widetilde{\Phi}^{(j)'}_{+,\textrm{P}}$ for $i \neq j$. From Eq. \eqref{eq:phi1a+p}  one finds $\widetilde{\Phi}^{(i)'\dag}_{+,\textrm{P}}\widetilde{\Phi}^{(j)'}_{+,\textrm{P}}=0$ for $i \neq j$. Therefore Eq. \eqref{eq:v6yp} is rewritten as
\be \label{eq:v6yp2}
\begin{split}
  V^{(6)}_{\mu\textrm{P}}&=\lambda^{(6)}_{\mu\textrm{P}} \left\{\left(\widetilde{\Phi}^{(1)'\dag}_{+,\textrm{P}}
                          \widetilde{\Phi}^{(1)'}_{+,\textrm{P}}\right)^2
                            +\left(\widetilde{\Phi}^{(2)'\dag}_{+,\textrm{P}}
                          \widetilde{\Phi}^{(2)'}_{+,\textrm{P}}\right)^2\right\},
\end{split}
\ee
and one finally arrives at the following identities,
\be \label{eq:P12dagP12P}
   \widetilde{\Phi}^{(1)'\dag}_{+,\textrm{P}}\widetilde{\Phi}^{(1)'}_{+,\textrm{P}}
   =\phi^{(1) \dag}_{\textrm{P}} \phi^{(1)}_{\textrm{P}} \; \,   \textrm{and} \; \,
   \widetilde{\Phi}^{(2)'\dag}_{+,\textrm{P}}\widetilde{\Phi}^{(2)'}_{+,\textrm{P}}
   =\phi^{(2) \dag}_{\textrm{P}} \phi^{(2)}_{\textrm{P}}.
\ee
After making use of Eqs. \eqref{eq:p1p} and \eqref{eq:p2p} and applying the following approach
\be
    \left(\frac{\Delta f'}{f'_i}\right)^2 \sim \left(\frac{\Delta f'}{f'_if'}\right)
    \sim 0,
\ee
where $f'_i = \frac{f_i}{\sqrt{\pi R}}, i=1,2.,f'^2=f'^2_1+f'^2_2, \Delta f'=f'_2-f'_1 $ and $H(H')=\frac{h}{\sqrt{\pi R}}\left(\frac{h'}{\sqrt{\pi R}} \right) $
a new field shall be defined as $H''=H'-H$ which has the order of SM VE$V$ i.e. $\mathcal{O}(\langle H''\rangle) \sim \mathcal{O}(v'')\sim \mathcal{O}(100$ Ge$V$) which will be clarified later [37]. These provide the following,
\begin{alignat}{2}\label{eq:p1dagp1}
     \phi^{(1)\dag}_{\textrm{P}} \phi^{(1)}_{\textrm{P}}
     &=f'^2_1 + 2\Delta f'f'_1+\frac{if'^2_2}{f'}\left(\begin{smallmatrix}
                                               \begin{smallmatrix}
                                                   0 & 0 \\
                                                   0 & 0 \\
                                                \end{smallmatrix}
                                                          & H'' \\
                                                 -H''^{\dag} & 0 \\
                                               \end{smallmatrix}\right)                               +\frac{f'^2_2}{f'^2}\left(
                                               \begin{smallmatrix}
                                                   HH^{\dag} & 0 \\
                                                   0 & (H'^{\dag}H') \\
                                                \end{smallmatrix}                                              \right),\\ \label{eq:p2dagp2}
     \phi^{(2)\dag}_{\textrm{P}} \phi^{(2)}_{\textrm{P}}
     &=f'^2_2 - 2\Delta f'f'_2+\frac{if'^2_1}{f'}\left(\begin{smallmatrix}
                                               \begin{smallmatrix}
                                                   0 & 0 \\
                                                   0 & 0 \\
                                                \end{smallmatrix}
                                                          & H'' \\
                                                 -H''^{\dag} & 0 \\
                                               \end{smallmatrix}\right)
                                               +\frac{f'^2_1}{f'^2}\left(
                                               \begin{smallmatrix}
                                                   H'H'^{\dag} & 0 \\
                                                   0 & (H^{\dag}H) \\
                                                \end{smallmatrix}                                              \right).
\end{alignat}

Besides the existing Baby Higgses field $H$ and $H'$ a new field $H''$ has emerged which will be clear later on as the SM-like Higgs. Having substituted Eqs. \eqref{eq:p1dagp1}, \eqref{eq:p2dagp2} into \eqref{eq:P12dagP12P} and further into \eqref{eq:v6yp2}, taking the mass terms and quartic terms, neglecting the constant field,  $V^{(6)}_{\mu\textrm{P}}$ now can be decomposed into 3 parts i.e potential of $H'', H'$ and $H$ and rewritten as
\be
    V^{(6)}_{\mu\textrm{P}} = V^{(6)}_{H''}+V^{(6)}_{H'}+V^{(6)}_{H}
\ee
where
\begin{alignat}{2}\label{eq:V6H}
    V^{(6)}_{H''} &= \lambda^{(6)}_{\mu\textrm{P}}\frac{f'^4_1 + f'^4_2}{f'^2}H''^{\dag}H'',\\ \label{eq:V6Hb}
    V^{(6)}_{H'}  &= \lambda^{(6)}_{\mu\textrm{P}}
                \left(\frac{f'^2_1f'^2_2}{f'^2}+\frac{(2 \Delta f' f'_1)f'^2_2}{f'^2} \right)H'^{\dag}H' + \lambda^{(6)}_{\mu\textrm{P}} \frac{f'^4_2}{f'^4}(H'^{\dag}H')^2,\\ \label{eq:V6Hc}
    V^{(6)}_{H} &= \lambda^{(6)}_{\mu\textrm{P}} \left(\frac{f'^2_1f'^2_2}{f'^2}-\frac {f'^2_1(2 \Delta f' f'_2)}{f'^2} \right)H^{\dag}H  + \lambda^{(6)}_{\mu\textrm{P}}  \frac{f'^4_1}{f'^4}(H^{\dag}H)^2.
\end{alignat}
All Eqs.\eqref{eq:V6H}, \eqref{eq:V6Hb} and \eqref{eq:V6Hc} have the mass terms now proving the broken shift symmetry. These show the important property of weakly-coupled potential and strongly indicate that $\lambda^{(6)}_{\mu \textrm{P}}$ is relatively small compared to $\lambda^{(6)}_{y}$. This also justifies that in weakly-coupled regime the interaction takes place at tree level which can produce mass.

If one assumes that $\Delta f' \ll f'_1 \sim f'_2 \sim f'$  in \susix \, scale then $V^{(6)}_{H'} \sim V^{(6)}_{H}$. In order to simplify further let's also assume $H \sim H'$ then one finds,
\be \label{eq:v6simp}
    V^{(6)}_{H'}+V^{(6)}_{H}=
    \lambda^{(6)}_{\mu \textrm{P}}
                \frac{2f'^2_1f'^2_2}{f'^2}H^{\dag}H+
                \lambda^{(6)}_{\mu \textrm{P}} \frac{f'^4_1+f'^4_2}{f'^4}(H^{\dag}H)^2\ \sim 2V^{(6)}_{H}
\ee
in which the radiative mass term of Little-like Higgs $(=\mu^2_{H}H^{\dag} H)$ and Little-like Higgs coupling are found to be,
\be \label{eq:m2H}
    \mu^2_{H} = \lambda^{(6)}_{\mu \textrm{P}} \frac{f'^2_1f'^2_2}{f'^2} \quad \textrm{and} \quad  \lambda_{\textrm{H}}=\lambda^{(6)}_{\mu \textrm{P}}\frac{f'^4_{1}+f'^4_2}{2f'^4}.
\ee
The mass of SM-like Higgs as produced by radiative symmetry breaking of \susix \, via Little-like Higgses from the  potential in Eq. \eqref{eq:v6simp}, and also Eq. \eqref{eq:m2H},
\be \label{eq:m2Ha}
    m^2_{H}=\frac{g^4}{16 \pi^2}\frac{\lambda^{(6)}_{\mu \textrm{P}}}{f'^2}(f'^2_1f'^2_2)\log\left\{\frac{\big(\Lambda^{\textrm{ZP}}_{(6)}\big)^2}{\mu^2_{H}} \right\},
\ee
where $\Lambda^{\textrm{ZP}}_{(6)}$ is the cut-off scale for zero-mode (perturbative) \susix \,  Baby Higgs theory above which \susix \, higher-mode perturbative and non-perturbative theory govern.

From Eq.\eqref{eq:m2H} with $\mu_{H} \sim \mathcal{O}(100 \textrm{Ge}V)$ and $f'_{i} \sim \mathcal{O}(1 \textrm{Te}V)$ one demonstrates that $\lambda^{(6)}_{\mu \textrm{P}}\ll\lambda^{(6)}_{y}=g^2_5$ or $\mathcal{O}(\lambda^{(6)}_{\mu \textrm{P}}) \sim  \mathcal{O}(\lambda_{H}) \sim  \mathcal{O}(10^{-2})$ to $\mathcal{O}(10^{-1})$ and justifies that  Little-like Higgses derived from \susix \, are weakly-coupled. Therefore zero mode-perturbative quartic coupling constant is much lower compared to $\lambda^{(6)}_{y}$ with a factor $\sim \mathcal{O}(10^{-2})$.

Another SM-like Higgs emerges from Eq. \eqref{eq:V6H} with the mass term $\mu^2_{H''}H''^{\dag}H''$ which gives
\be \label{eq:m2H'a}
    m^2_{H''}=\frac{g^4}{16 \pi^2}\frac{\lambda^{(6)}_{\mu \textrm{P}}}{f'^2}(f'^4_1+f'^4_2)\log\left\{ \frac{\big(\Lambda^{\textrm{ZP}}_{(6)}\big)^2}{\mu^2_{H''}}\right\},
\ee
with $\mu_{H''} \sim \mathcal{O}(100 \textrm{Ge}V)$.

The factor $f'^2_1f'^2_2$ in Eq. \eqref{eq:m2Ha} will reach maximum value at $f'_1=f'_2$ if $f'^2_1+f'^2_2$ is set constant and gives accordingly the interval $m_{H}<m_{1}$. On the other hand $(f'^4_1+f'^4_2)$ in Eq. \eqref{eq:m2H'a} will obtain its minimum value at $f'_1=f'_2$ if $(f'^2_1f'^2_2)$ is set constant and provides the interval $m_{H''}>m_{2}$. One can conclude that there exist the exclusion area for Higgs mass $m_{1}<m_{\textrm{Higgs}}(\textrm{excluded}) <m_{2}$.

Eq. \eqref{eq:m2Ha} gives the mass of light SM-like Higgs while Eq. \eqref{eq:m2H'a} clearly provides intermediate SM-like Higgs since $f'^4_1+f'^4_2 > f'^2_1f'^2_2$.

Two Little-like Higgses, $H$ and $H'$, of Eq. \eqref{eq:m2Ha} and the new Higgs, $H''$, of Eq. \eqref{eq:m2H'a} clearly form triplets of Higgses and when they are put in \susix \, multiplet one find two sextets $6_{H}=\begin{pmatrix}
                                  H & H' & H'' \\
                               \end{pmatrix}^{T}$ and one decapentuplet
$15_{H}=
 \left(\begin{smallmatrix}
                           (0)_{3 \times 3} & \left( \begin{smallmatrix}
                                               \begin{smallmatrix}
                                                   0 & 0 \\
                                                   0 & 0 \\
                                                \end{smallmatrix}
                                                          & (H) \\
                                                 (H'^{\dag}) & 0 \\
                                               \end{smallmatrix} \right)
 \\
                           \left(\begin{smallmatrix}
                                               \begin{smallmatrix}
                                                   0 & 0 \\
                                                   0 & 0 \\
                                                \end{smallmatrix}
                                                          & (H') \\
                                                 (H^{\dag}) & 0 \\
                                               \end{smallmatrix}\right)  & \left(\begin{smallmatrix}
                                               \begin{smallmatrix}
                                                   0 & 0 \\
                                                   0 & 0 \\
                                                \end{smallmatrix}
                                                          & (H'') \\
                                                 (H''^{\dag}) & 0 \\
                                               \end{smallmatrix}\right) \\
                         \end{smallmatrix}\right)$ [2,20]. With their hermitian conjugate they give totally 18 (Little-like) Higgses to be eaten by the will-be-massive gauge bosons and bring about symmetry breaking \susix$\rightarrow$\suthr$\times$\suthr$\times$\uone.

This is basically the area of weakly-coupled \susix \, Baby Higgs i.e. $\Lambda_{(3)}<\textrm{WC}<\Lambda^{\textrm{NP}}_{(6)}$ where GUT based on \susix \, symmetry also lives and governs. (Appendix A.1)

\section{Phenomenological Aspects}
\subsection{Unification of gauge coupling constant}

The three couplings $g_{c}, g_{w}$ and $g_{em}$ in the brane $y=0$ due to symmetry \suthrc$\times$\sutwol$\times$\uoney \, run linearly in logarithmic scale even into the near-brane area below the compactification scale $M_c$. The reason is clear for below $M_c$ the 4D-property  is dominant $(\Lambda^{4\textrm{D}}_{(6)} \sim M_{c})$. Above $M_c$ with more dominant  5D-property the couplings   shall  curve logarithmically  until $M_{\ast} \sim 10^{9}$ Te$V$  with value close to $0.6-0.7$ and continue to be power law until approaching $g_{5} \sim 4 \pi$ at $\Lambda^{5\textrm{D}}_{(6)}$.

Following ref [14] and [32] one has the following plot for $M_{c} \sim 10^{10}$ Ge$V$. Now it is clear for 4D \susix \,  weakly-coupled GUT based on zero mode with $\Lambda_{(6)}^{\textrm{ZP}} \sim 1000$ Te$V$ the coupling constants reach value $g_{4} \sim 0.7$. More elaborate discussion is given in Appendix B.

\subsection{Higgs spectrum and masses}
Latest LHC data on Higgs mass excluded region lie in the interval 145-466 Ge$V$. On the other hand the near-brane weakly-coupled \susix \,  Baby Higgses do provide beyond-SM Higgses of light and  heavy types. If one sets $f'_1 \sim 4.0$ Te$V$, and $f'_2 \sim 5.0$ Te$V$ $g_{4}=0.7$ and $ \mu_{H} \sim \mu_{H'} \sim \mu_{H''} \sim 100$ Ge$V$ and $\lambda^{(6)}_{\mu \textrm{P}}=0.10$ with cut-off scale $\Lambda_{(6)}^{\textrm{ZP}} \sim 1000$ Te$V$ then masses of Higgses are $108$ Ge$V$ and $160$ Ge$V$ for $H(=H')$ and $H''$ which are exactly (light) SM Higgs for $H$ and excluded intermediate Higgs for $H''$. This proves that electroweak scale has been increased up $10$ Te$V$. But if $f'_i, i=1,2$ is set at $O (10$ Te$V$) such as $f'_1= 16$  Te$V$ and $f'_2= 20 $  Te$V$, and other parameters remain the same then one obtains Higgs masses of $432$ Ge$V$ and $640$ Ge$V$ for excluded light Higgs for  $H$ and intermediate Higgs for $H''$ which is well under unitary constraint of $\sim 700$ Ge$V$. These  provide the intermediate Higgs boson. If one adds $\delta_{m_{H}} \sim 10$-$20$ Ge$V$ due to radiative correction the light and intermediate Higgses lie in the most preferred mass interval of Higgs boson. This result confirms that $H$ is light with maximum $f'_i=5.92$ Te$V$ and $H''$ is intermediate (heavy) Higgs bosons with minimum $f'_i=13.45$ Te$V$ and proves that the excluded region corresponds to VE$V$  interval $(5.9-13.45)$ Te$V$.

\subsection{Proton decay}
Another important phenomenological constraint is proton decay. Fortunately, proton decay in the current model can be kept long enough to fulfill the experimental bound [20]. One of the reasons the leptoquark like interaction at tree level is not allowed at all. There are actually two reasons for this behavior, i.e. at the \suthr \,  scale, the \suthr \, triplets containing
quarks and leptons generated from the \susix \, sextet are completely separated. Obviously there is no tree level interaction between both of them, and, at the \susix \, scale, as explained in [35] and shown by the sub-generators $\lambda_{C^{(1,2)}}$ in [35],
all charges of leptoquark-like gauge bosons in the model are integer. This disallows tree level quark and lepton interaction which should require gauge bosons with fractional charges.

Possible decay due to baryon and lepton number violating dimension-6 operators generated by quantum correction is suppressed by $1/M^{2}_{(5\textrm{D})}$ where $M_{(5\textrm{D})} \sim \Lambda^{\textrm{NP}}_{(6)} \sim 10^{8}$ Ge$V$ with $M_{(5\textrm{D})}$ is 5D-origin GUT scale (Appendix A.2) which is equivalent to the suppression factor of dimension-5 operator in conventional 4D GUT scale $1/M_{(4\textrm{D})}$ with $M_{(4\textrm{D})} \sim 10^{16}$ Ge$V$. This confirms the same proton stability as known in conventional 4D GUT. Alternative scheme for protecting proton lifetime can also be provided by this model through localizing wave function on the brane $y=0$ for SM particles [32,36] with thickness $L=(M^{*})^{-1}$ separating baryon from lepton or just putting baryon in the brane and lepton in the near-brane at the proximity beyond separating distance $L$. Proton decay starts to take place at $M^{*}$ as low as $1.0$ Te$V$ [32,36] which is very much lower than $M^{*}=10^{12}$ Ge$V$ in this model (Appendix A.2).

Therefore, roughly speaking the model should predict the proton life time close to the SM's one. Of course, more investigation should be done properly in a separate work.

\section{Conclusion}
The \susix-origin Little (Baby) Higgses as the by-product of Scherk-Schwarz mechanisms and orbifold $S^1/Z_2$ breaking with duality in trivial and pseudo non-trivial manners have decoupled from \susix \, GUT and been replaced immediately by \susix \, will-be-SimplestLittleHiggses and Baby Higgs. This approach is realized mainly by utilizing  the zero mode terms, so that it shows up as weakly-coupled  in contrast with the strongly-coupled \susix-origin Little (Baby)  Higgses. Now, it is clear that the dimensional deconstruction and symmetry breaking of 5D \susix \, happen almost at once due to duality condition.

This brings about the reduction (elimination) of higher modes in trivial manner and the splitting of parity in pseudo non-trivial manner. Consequently as final product, strongly-coupled \susix-origin Little (Baby) Higgses transform itself into final \suthr \, Simplest Little Higgses which are  strongly-coupled and \suthr \, Little-like  Higgses which are weakly coupled as will be discussed further in a separate paper [37]. These \susix \, triplet-contained Higgses, by means of radiative symmetry breaking, break \susix \, symmetry (and electroweak symmetry later on) and produce SM-like Intermediate Higgs bosons with masses $\sim 470$ - $650$ Ge$V$s for VE$V$s $14.0-20.0$ Te$V$ and SM Light Higgs with masses $100-110$ Ge$V$ for VE$V$s $4.0-5.0$ Te$V$ both at $g_{4} \sim 0.7$. Unification scale is  of $10^{12}$ Ge$V$ and compactification scale at $R^{-1}\sim 10^{10}$ Ge$V$.
Extra dimension is of the order of $10^{-24}$ cm. Some observables are already within reach of LHC in the current time.

\section*{Acknowledgements}
AH would like to thank PT Enerfra Septadaya Prima for administrative support while FPZ to ITB for research fund (fiscal year 2012).

\appendix

\section{}

\subsection{Scales for Perturbative \susix \, GUT}
\begin{figure}[h]
\centering
\includegraphics[width=9.5cm, height=5.5cm, keepaspectratio=true]{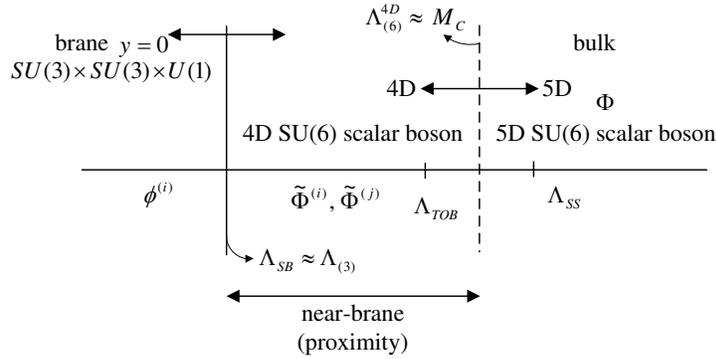}
\caption{\label{Near-brane-4DSU(6)}\small{Near-brane 4D-\susix \, scalar boson cut-off scale.}}
\end{figure}

Let us define Trivial Orbifold Breaking (TOB), or pseudo Non-trivial Breaking, at energy level $\Lambda_{\textrm{TOB}}$ while  (Radiative) Symmetry Breaking (SB) at energy level $\Lambda_{\textrm{SB}}$ which is assumed to be a bit higher than $\Lambda_{\textrm{3}}$ the cut-off scale of $\textrm{4D-\suthr}\times\textrm{\suthr}\times\textrm{\uone}$ theory. We also define $\Lambda_{\textrm{SS}}$, Scherk-Schwarz (SS) symmetry breaking level which is taken to be higher than $\Lambda_{\textrm{TOB}}$. For clarity one can draw the following energy scale and plot the valid interval for near-brane (proximity) area which represents 4D \susix \,  area obtained from Sec. 2.3.1, as shown in Fig. \ref{Near-brane-4DSU(6)}.
\\
Later on near-brane interval will be divided into perturbative \susix \,  and non-perturbative \susix \, theory. We define compactification scale $M_{c}$ which is slightly higher than $\Lambda^{4\textrm{D}}_{(6)}$, the cut-off scale of 4D-\susix \, theory, in the following equation
\be
        M_{c}=\frac{1}{R}
\ee
where $R$ is the compactification radius and unification scale (of gauge coupling) $M_{\ast}$ with the following constraint [31,32]
\be
        M_{\ast}R \sim \mathcal{O}(100).
\ee
For our purpose we may take $M_{\ast} \sim 100 M_{c}$ [37-38].

\begin{figure}[h]
\centering
\includegraphics[width=9.5cm, height=5.5cm, keepaspectratio=true]{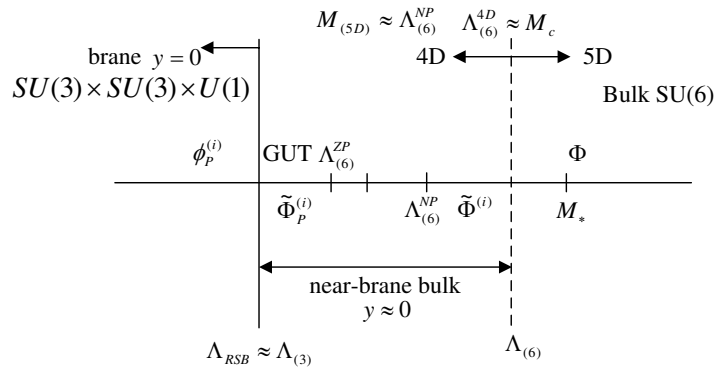}
\caption{\susix \, weakly-coupled GUT and \susix-origin strongly-coupled  Little (Baby) Higgs.}
\label{weakly-coupled}
\end{figure}

Following ansatz is taken: below $M_{c}$ the near-brane area shows dominant 4D-property while above $M_{c}$ the near-brane (now becoming bulk) has dominant 5D-property. In this way one finds perturbative approach is valid below $M_{c}$ and results in so-called weakly-coupled \susix \, Baby Higgs which is suitable for \susix \, GUT from $\Lambda_{3}$ up to $\Lambda_{(6)}^{\textrm{NP}} \sim M_{(5\textrm{D})}$, the cut-off scale of 5D-origin \susix \, GUT theory. On the other hand non-perturbative approach must be taken  above $\Lambda_{(6)}^{\textrm{NP}}$ due to more (and more) Kaluza-Klein states causing more dominant 5D-property. Higher modes at maximum  will bring about the strongly-coupled property of the \susix-origin Little (Baby) Higgs, a by-product of this model, serving as the origin of \susix \, Baby Higgs. It is clear later on that the interval $(\Lambda_{3}, \Lambda_{(6)}^{\textrm{NP}})$ does consist of $\Lambda_{(6)}^{\textrm{ZP}}, \Lambda_{(6)}^{\textrm{FP}},\cdots$ etc, as the cut-off scale of zero mode-based \susix \, GUT and first mode-based \susix \, GUT, etc., as shown in Fig. \ref{weakly-coupled}.
\\
Finally the zero mode-based weakly-coupled \susix \, Baby Higgses are given in the Fig. \ref{Zero mode and first mode} below.
\begin{figure}[h]
\centering
\includegraphics[width=9.5cm, height=5.5cm, keepaspectratio=true]{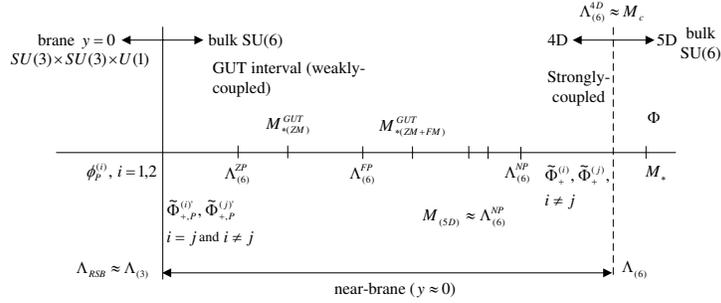}
\caption{Zero mode- and first mode- based weakly-coupled  \susix \, Baby Higgs in the near-brane ($y \sim 0$).}
\label{Zero mode and first mode}
\end{figure}
\subsection{Order of Estimate}
Planck scale is reduced significantly in 5D and so is GUT scale following the formula below,
\be
    \Big[M_{P}^{(5D)}\Big]^{D-2} = \Big[M_{P}^{(4D)}\Big]^{2}/R^{\delta}
\ee
where
\be \nonumber
    \begin{array}{lll}
    D      &= 4+\delta = 5,  &\hbox{is space-time dimension;} \\
    \delta &=1,  &\hbox{is number of extra-dimension;} \\
    \frac{1}{R} &=M_{c} &\hbox{is compactification scale.}
  \end{array}
 \ee
With $M_{P}^{(4D)} \sim 10^{19}$ Ge$V$ and $M_{c}$ is assumed to be $10^{10}$ Ge$V$ one finds the five-dimensional Planck scale, $(M_{P}^{(5D)})$ in the order of $ \sim 10^{16}$ Ge$V$ or $10^{-3} M_{P}^{(4D)}$. This brings us to the 5D-\susix GUT unification scale of $10^{12}$ Ge$V (M_{\ast})$ and compactification scale $M_{c} \sim 10^{-2} M_{\ast} \sim  10^{10}$ Ge$V$  which has justified the previous assumption. This in turn establishes the limit for the \susix \, GUT perturbative cut-off scale, $M_{(5\textrm{D})}=10^{8} \textrm{Ge}V \sim \Lambda_{(6)}^{\textrm{NP}} < \Lambda_{(6)}^{4\textrm{D}} \sim 10^{7} $ Te$V$ and minimal cut-off scale of 5D \susix theory, $\Lambda^{5\textrm{D}}_{6}$, as $3\times 10^{10}$ Te$V$  due to $(\Lambda_{6}/M_{\ast}) |_{\delta=1} < 30$ [37,38].

\section{Unification of Gauge Coupling Constant}
In general the logarithmic form of running coupling constant in 4D is changed into a power law due to the effect of extra dimension in accordance with the following formula [37],
\be
\begin{split}
    \alpha_{i}^{-1}(\Lambda) &= \alpha_{i}^{-1}(\mu)-\frac{b_{i}
              -\widetilde{b_{i}}}{2 \pi}\ln\frac{\Lambda}{\mu} -\frac{\widetilde{b_{i}}}{4 \pi} \int^{r \mu^{-2}}_{r \Lambda^{-2}}\frac{dt}{t} \left\{v_{3}\left(\frac{it}{\pi R^{2}}\right) \right\}^{\delta}
\end{split}
\ee
where Jacobi theta-function $v_{3}(\tau) \equiv \sum^{+\infty}_{n=-\infty} \exp(\pi i \tau n^{2})$ reflects the sum over K-K states, $\widetilde{b_{i}}$  are the beta-function coefficients,  $r \equiv \pi (X _{\delta})^{-2/\delta}$ with $X _{\delta}= 2 \pi^{\delta/2}/\delta\Gamma (\delta/2)$ as overall normalization.

\begin{figure}[h]
\centering
\includegraphics[width=6.5cm, height=4.5cm, keepaspectratio=true]{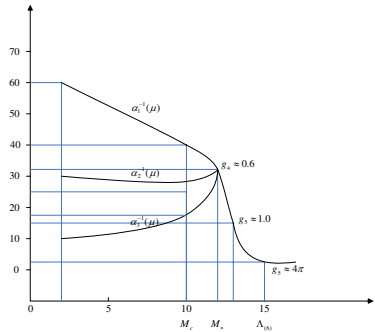}
\caption{\label{Unification-of-gauge}\small{Unification of gauge couplings with a single extra dimension of radius $R^{-1} \approx 10^{10}$ Ge$V$.}}
\end{figure}

This can be modified into the following
\be
    \alpha_{i}^{-1}(\Lambda) = \alpha_{i}^{-1}(R^{-1})-\frac{b_{i}
              -\widetilde{b_{i}}}{2 \pi}\ln (\Lambda R) -\frac{\widetilde{b_{i}}X_{\delta}}{2 \pi \delta} \Big[(\Lambda  R)^{\delta}-1\Big]
\ee
where the cut-off scale $\Lambda \gg \frac{1}{R}$ or \susix \, cut off scale $\Lambda^{5\textrm{D}}_{(6)}$ is set in such a way $\Lambda^{5\textrm{D}}_{(6)} \gg M_{c}$. Above $M_{c}$ the last term of (B.2) becomes more and more influential. This gives rise to power-law evolution which is basically valid until the limit $\Lambda R \approx 1$ or $\Lambda \approx R^{-1}=  M_{c}$ .
Here we denote $\Lambda$ as $\Lambda_{(6)}^{5\textrm{D}}$ which is the cut-off scale of 5D-\susix \,  starting from $\Lambda_{(6)}^{\textrm{5D}}$ downward to where $\Lambda_{(6)}^{4\textrm{D}} \approx M_{c}$. Below $M_{c}$ in GUT area, perturbative approach becomes reliable which basically provides the weakly-coupled area down to symmetry breaking level $\Lambda_{(3)}$. Result is plotted, based on [14] and [38] as in   in  Fig. \ref{Unification-of-gauge}.
\\

\end{document}